\documentclass[conference,letterpaper,romanappendices]{IEEEtran}

\usepackage[utf8]{inputenc} 
\usepackage[T1]{fontenc}
\usepackage{url}
\usepackage{ifthen}
\usepackage{cite}

\usepackage{graphicx}
\usepackage{hyperref}
\usepackage{amsmath}
\usepackage{amssymb}
\usepackage{bm}
\usepackage{bbm}
\usepackage{algorithmicx}
\usepackage{algorithm}
\usepackage{algpseudocode}
\usepackage{array}
\usepackage{booktabs}
\usepackage{multirow}
\usepackage{makecell}
\usepackage{mathtools}
\usepackage{amsthm}
\usepackage{etoolbox}

\newtheorem{definition}{Definition}
\newtheorem{theorem}{Theorem}
\newtheorem{remark}{Remark}
\newtheorem{lemma}{Lemma}
\newtheorem{corollary}{Corollary}

\newcommand{\F}{\mathcal{F}}
\newcommand{\E}{\mathbb{E}}

\newcommand{\indicator}{\mathbbm{1}}
\newcommand{\Y}{\mathcal{Y}}
\newcommand{\X}{\mathcal{X}}
\newcommand{\St}{\mathcal{S}}
\newcommand{\thetai}{i}

\DeclareMathOperator*{\argmax}{arg\,max}

\begin{document}
\title{Finite-Blocklength Performance of Sequential Transmission over BSC with Noiseless Feedback}


\author{\IEEEauthorblockN{Hengjie~Yang and Richard~D.~Wesel}
\IEEEauthorblockA{
Department of Electrical and Computer Engineering\\
University of California, Los Angeles, Los Angeles, CA 90095, USA\\
Email: \{hengjie.yang, wesel\}@ucla.edu}
}


\maketitle

\begin{abstract}
In this paper, we consider the problem of sequential transmission over the binary symmetric channel (BSC) with full, noiseless feedback. Naghshvar \emph{et al.} proposed a one-phase encoding scheme, for which we refer to as the small-enough difference (SED) encoder, which can achieve capacity and Burnashev's optimal error exponent for symmetric binary-input channels. They also provided a non-asymptotic upper bound on the average blocklength, which implies an achievability bound on rates. However, their achievability bound is loose compared to the simulated performance of SED encoder, and even lies beneath Polyanskiy's achievability bound of a system limited to stop feedback. This paper significantly tightens the achievability bound by using a Markovian analysis that leverages both the submartingale and Markov properties of the transmitted message.  Our new non-asymptotic lower bound on achievable rate lies above Polyanskiy's bound and is close to the actual performance of the SED encoder over the BSC.
\end{abstract}

\section{Introduction}

Feedback does not increase the capacity of memoryless channels \cite{Shannon1956}, but it can significantly reduce the complexity of communication and the probability of error, 
provided that variable-length feedback (VLF) codes are allowed. In his seminal paper, Burnashev \cite{Burnashev1976} first proposed a conceptually important two-phase transmission scheme for any discrete memoryless channel (DMC) with noiseless feedback. The first phase is called the \emph{communication phase}, in which the transmitter seeks to increase the decoder's belief about the transmitted message by improving its posterior to above $1/2$. The second phase is called the \emph{confirmation phase}, in which the transmitter seeks to increase the posterior of the most likely message identified from the communication phase to above a target threshold, at which it can be reliably decoded. Burnashev's two-phase encoding scheme yields the optimal error exponent for the DMC with noiseless feedback. 

For the binary symmetric channel (BSC) with noiseless feedback, Horstein \cite{Horstein1963} first proposed a simple, elegant transmission scheme that achieves the capacity of the BSC. However, a rigorous proof of its capacity-achieving property remained elusive until the work of Shayevitz and Feder \cite{Shayevitz2011} which generalizes Horstein's idea to the concept of  posterior matching. Since Horstein's work, several authors have constructed schemes to achieve the capacity or the optimal error exponent of BSC with noiseless feedback; see \cite{Schalkwijk1971,Schalkwijk1973,Tchamkerten2002,Tchamkerten2006,Naghshvar2012}.

Recently, attention has shifted from the asymptotic regime, which focused on long average blocklength at a fixed rate and probability of error, to the finite-blocklength regime. Polyanskiy \emph{et al.} \cite{Polyanskiy2010,Polyanskiy2011} first showed that variable-length coding with noiseless feedback can provide a significant advantage in achievable rate over fixed-length codes without feedback.  In their analysis, a simple stop feedback scheme is enough to obtain an achievable rate larger than that of a fixed-length code without feedback. For practical communications, Williamson \emph{et al.} \cite{Williamson2015} investigated how coding techniques using feedback can approach capacity as a function of average blocklength.

For symmetric binary-input channels with noiseless feedback, Naghshvar, Javidi and Wigger \cite{Naghshvar2012,Naghshvar2015} proposed a deterministic encoding scheme, which we refer to as the \emph{small-enough difference} (SED) encoder, which attains capacity and Burnashev's optimal error exponent. They also gave a non-asymptotic upper bound on the average blocklength of the SED encoder. However, in the case of BSC, their bound corresponds to a lower bound on achievable rate that lies beneath Polyanskiy's lower bound on the achievable rate of a system limited to stop feedback. A system such as the SED encoder that exploits full noiseless feedback should provide a higher rate than a system limited to stop feedback.

In this paper, we seek a tightened upper bound on average blocklength of sequential transmission over BSC with full, noiseless feedback.  The bounds of \cite{Naghshvar2012,Naghshvar2015} were derived by synthesizing a delicate new submartingale from two submartingales that characterize the fundamental behavior of the transmitted message. In fact, this general proof technique dates back to the work of Burnashev and Zigangirov \cite{Burnashev1975} and was later generalized by Naghshvar \emph{et al.}\cite{Naghshvar2012,Naghshvar2015}. This sophisticated analysis succeeds in establishing a non-asymptotic upper bound, but it does not reveal the fundamental mechanism that produces the constant term in the bound.

Following the SED encoder in \cite{Naghshvar2015}, we present a Markovian analysis that leverages the submartingale results of Naghshvar \emph{et al.}\cite{Naghshvar2012,Naghshvar2015} and the Markov structure of the the transmitted message during its confirmation phase. This enables us to significantly tighten the upper bound on average blocklength and to gain a deep understanding of the constant term in the bound. Specifically, we will apply a time of first-passage analysis on the Markov chain formed by the transmitted message in the confirmation phase, which fully accounts for the times when the transmitted message ``falls back'' from the confirmation phase to the communication phase. Our analysis reveals that the constant term mainly results from the differential time spent in the ``fallback'' stage.

The organization of this paper is as follows. In Sec. \ref{sec: problem setup} formulates the problem of sequential transmission over DMC with noiseless feedback and introduce Naghshvar \emph{et al.}'s scheme. Sec. \ref{sec: markovian analysis} reviews some previous results, and presents our main result and the proof using our Markovian analysis. Sec. \ref{sec: simulation} demonstrates the simulated performance of the SED encoder and compares our results with the previous achievability bound by Polyanskiy and a bound resulted from a lemma of Naghshvar \emph{et al.}

\begin{figure}[t]
\centering
\includegraphics[width=0.45\textwidth]{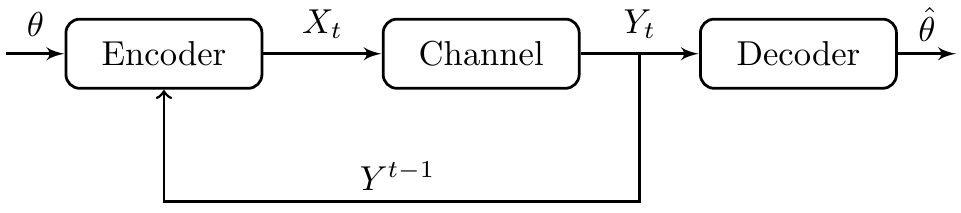}
\caption{System diagram of a DMC with full, noiseless feedback.}
\label{fig: system model}
\end{figure}

\section{Problem Setup}
\label{sec: problem setup}

Consider the problem of sequential transmission (or variable-length coding) over a DMC with full, noiseless feedback as depicted in Fig. \ref{fig: system model}. The DMC is described by the finite input set $\mathcal{X}$, finite output set $\mathcal{Y}$, and a collection of conditional probabilities $P(Y|X)$. The Shannon capacity of the DMC is given by
\begin{align}
C=\max_{P_X} I(X;Y),
\end{align}
where $P_X$ denotes the probability distribution over finite set $\X$. Let $C_1$ be the maximal Kullback-Leibler (KL) divergence between the conditional output distributions,
\begin{align}
C_1=\max_{x,x'\in\X}D\big(P(Y|X=x)\| P(Y|X=x') \big).
\end{align}
We also denote
\begin{align}
C_2=\max_{y\in\Y}\log\frac{\max_{x\in\X}P(Y=y|X=x)}{\min_{x\in\X}P(Y=y|X=x)}.
\end{align}
All logarithms in this paper are base $2$. We assume $C,C_1,C_2$ are positive and finite. It can be easily shown that $0<C\le C_1\le C_2<\infty$. For BSC$(p)$ with crossover probability $0<p<1/2$, letting $q=1-p$, we have
\begin{align}
C=&1-H(p)\\
C_1=&p\log\frac{p}{q}+q\log\frac{q}{p}\\
C_2=&\log\frac{q}{p}.
\end{align}

Let $\theta$ be the transmitted message uniformly drawn from the message set $\Omega=\{1,2,\dots, M\}$. The total transmission time (or the number of channel uses, or blocklength) $\tau$ is a random variable that is governed by some stopping rule as a function of the observed channel outputs. Thanks to the noiseless, feedback channel, the transmitter is also informed of the channel outputs and thus the stopping time.

The transmitter wishes to communicate $\theta$ to the receiver. To this end, it produces channel inputs $X_t$ for $t=1,2,\dots,\tau$ as a function of $\theta$ and past channel outputs $Y^{t-1}=(Y_1,Y_2,\dots,Y_{t-1})$, available to the transmitter through the full, noiseless feedback channel. Namely,
\begin{align}
X_t=e_t(\theta,Y^{t-1}),\quad t=1,\dots,\tau, \label{eq: encoding rule}
\end{align}
for some encoding function $e_t: \Omega\times \Y^{t-1}\to \X$.

After observing $\tau$ channel outputs $Y_1,Y_2,\dots,Y_{\tau}$, the receiver makes a final estimate $\hat{\theta}$ of the transmitted message as a function of $Y^{\tau}$, i.e.,
\begin{align}
\hat{\theta}=d(Y^{\tau}), \label{eq: decoding rule}
\end{align}
for some decoding function $d: \Y^\tau\to \Omega$. 

The probability of error of the scheme is given by
\begin{align}
P_e\triangleq \Pr\{\hat{\theta}\ne \theta\}.
\end{align}

For a fixed DMC and for a given $\epsilon>0$, the goal is to find encoding and decoding rules described in \eqref{eq: encoding rule}, \eqref{eq: decoding rule}, and a stopping time $\tau$ such that $P_e\le \epsilon$ and the average blocklength $\E[\tau]$ is minimized.

As noted in \cite{Naghshvar2015}, a sufficient statistic  of $Y^{t-1}$ for $\theta$ is the \emph{belief state} of the receiver,
\begin{align}
\bm{\rho}(t)=[\rho_1(t),\rho_2(t),\dots,\rho_M(t)],\quad t=0,1,2,\dots,\tau,
\end{align}
where for each $i\in\Omega$, $\rho_i(t)=\Pr\{\theta=i|Y^t\}$ for $t\ge1$, and $Y^0=\emptyset$. The receiver's initial belief of $\theta=i$ is $\rho_i(0)=\Pr\{\theta=i\}=1/M$.  According to Bayes' rule, upon receiving $y_t$, $\rho_i(t)$ can be updated by
\begin{align}
\rho_i(t)=\frac{\rho_i(t-1)P(Y=y_t|X=e_{t}(i,Y^{t-1}))}{\sum_{j\in\Omega}\rho_j(t-1)P(Y=y_t|X=e_{t}(j,Y^{t-1}))}. \label{eq: Bayes's rule}
\end{align}
Thanks to the noiseless feedback, the transmitter will be informed of $y_t$ at $t+1$ and thus can calculate the same $\bm{\rho}(t)$. The stopping time $\tau$ and decoding rule considered in \cite{Naghshvar2015} are given by
\begin{align}
\tau=&\min\{t: \max_{i\in\Omega}\rho_i(t)\ge1-\epsilon\}\label{eq: stopping rule}\\
\hat{\theta}=&\argmax_{i\in\Omega}\rho_i(\tau).\label{eq: estimate message}
\end{align}
Clearly, with the above scheme, the probability of error meets the desired constraint, i.e.,
\begin{align}
P_e=\E[1-\max_{i\in\Omega}\rho_i(\tau)]\le\epsilon.
\end{align}

For any DMC, Naghshvar \emph{et al.} \cite{Naghshvar2012,Naghshvar2015} proposed an encoder, which we refer to as the \emph{small-enough difference} (SED) encoder, for symmetric binary-input channels (thus also for the BSC). This encoder is implemented using a partitioning algorithm, which, after calculating $\bm{\rho}(t-1)$, partitions $\Omega$ into two subsets $S_0(t-1)$ and $S_1(t-1)$ such that
\begin{align}
0\le\sum_{\mathclap{i\in S_0(t-1)}}\ \rho_i(t-1)-\sum_{\mathclap{i\in S_1(t-1)}}\ \rho_i(t-1)<\min_{i\in S_0(t-1)}\rho(t-1). \label{eq: MaxEJS encoder}
\end{align}
Then, $X_t=0$ if $\theta\in S_0(t-1)$ and $X_t=1$ otherwise.

With the stopping time in \eqref{eq: stopping rule} and the SED encoder in \eqref{eq: MaxEJS encoder}, Naghshvar \emph{et al.} proved the following non-asymptotic upper bound on $\E[\tau]$ via a delicate submartingale synthesis.
\begin{theorem}[Remark 7, \cite{Naghshvar2015}]\label{theorem: Naghshvar et al}
The proposed scheme described in \eqref{eq: stopping rule}, \eqref{eq: estimate message}, and \eqref{eq: MaxEJS encoder}, for symmetric binary-input channels satisfies,
\begin{align}
\E[\tau]\le\frac{\log M+\log\log\frac{M}{\epsilon}}{C}+\frac{\log\frac{1}{\epsilon}+1}{C_1}+\frac{96\cdot2^{2C_2}}{CC_1}. \label{eq: Naghshvar et al}
\end{align}
\end{theorem}

\begin{remark}
We make several remarks regarding Theorem \ref{theorem: Naghshvar et al}. First, the proof of Theorem \ref{theorem: Naghshvar et al} involves Doob's optional stopping theorem \cite{Williams:1991} and a delicate construction of a new submartingale that combines two submartingales similar to that in Lemma \ref{lemma: behavior of transmitted msg}. We refer interested readers to the Appendix of \cite{Naghshvar2015} for complete proof details. In fact, this general proof technique dates back to the work of Burnashev and Zigangirov \cite{Burnashev1975} and was later generalized by Naghshvar et al. \cite{Naghshvar2015}. However, such sophisticated analysis leaves readers with little insight about the constant term in \eqref{eq: Naghshvar et al}. Second, our simulations will show that, for the BSC$(0.05)$, the achievability bound from Theorem \ref{theorem: Naghshvar et al} is loose enough that it does not capture the actual performance of the SED encoder. This bound even falls below Polyanskiy's VLF lower bound that characterizes the achievable rate of a system limited to stop feedback.
\end{remark}

\section{The Markovian Analysis on Average Blocklengths}
\label{sec: markovian analysis}

In this section, we consider the problem of sequential transmission (or variable-length coding) over BSC with full, noiseless feedback. Specifically, we follow Naghshvar \emph{et al.}'s framework described in Sec. \ref{sec: problem setup}, i.e., the stopping time in \eqref{eq: stopping rule}, the decoding rule in \eqref{eq: estimate message}, and the SED encoder in \eqref{eq: MaxEJS encoder}. Our analysis focuses on BSC$(p)$ with crossover probability $0<p<1/2$.

Unlike the proof technique of Theorem \ref{theorem: Naghshvar et al}, we propose a \emph{Markovian analysis}. First, we decompose the process into a communication phase and a confirmation phase that also takes into account the fallback of the transmitted message, i.e., the time when the transmitted message falls back from the confirmation phase to the communication phase and then returns to the confirmation phase. Next, we utilize submartingale results for the communication phase, but exploit the Markov structure of the confirmation phase to perform a time-of-first passage analysis.   The constant term in the time of first-passage analysis explicitly captures the penalty  of falling back, and this same constant term appears in our final bound. Eventually, our analysis yields the following tight upper bound on $\E[\tau]$.
\begin{theorem}\label{theorem: main results}
The proposed scheme described in \eqref{eq: stopping rule}, \eqref{eq: estimate message}, and \eqref{eq: MaxEJS encoder} for the BSC$(p)$, $0<p<1/2$, satisfies
\begin{align}
\E[\tau]\le\frac{\log M}{C}{+}\frac{\left\lceil \frac{\log\frac{1-\epsilon}{\epsilon}}{C_2} \right\rceil C_2}{C_1}{+}\frac{pC_2}{C_1}\left(\frac{C_1+C_2}{C}-\frac{C_2}{C_1}\right){+}\frac{C_1}{C}.\label{eq: main result}
\end{align}
\end{theorem}

For brevity, throughout Sec. \ref{sec: markovian analysis}, denote by $\theta=i\in \Omega$ the transmitted message unless otherwise specified.

\subsection{Previous Results of Naghshvar et al. and Polyanskiy}
We first review several key results Naghshvar \emph{et al.} demonstrated in \cite{Naghshvar2012} and \cite{Naghshvar2015} and Polyanskiy's VLF upper bound derived by Williamson \emph{et al.} \cite{Williamson2015}. 

For shorthand notation, let $\rho_{i}(t)$ denote the posterior of the transmitted message $\theta = i\in\Omega$. The log-likelihood ratio of $\theta = i$ is denoted 
\begin{align}
U_{\thetai}(t)=\log\frac{\rho_{\thetai}(t)}{1-\rho_{\thetai}(t)}.
\end{align}
For a given $\epsilon>0$, define the genie-aided stopping time $\tau_{\thetai}(\epsilon)$ of $\theta = i$ as
\begin{align}
\tau_{\thetai}(\epsilon)=\min\{t: \rho_{\thetai}(t)\ge1-\epsilon\}. \label{eq: genie-aided decoder}
\end{align}

With the SED encoding rule described in \eqref{eq: MaxEJS encoder}, Naghshvar \emph{et al.} proved that $\{U_{\thetai}(t)\}_{t=0}^{\infty}$ forms a submartingale.

\begin{lemma}[Naghshvar \emph{et al.}, \cite{Naghshvar2012}]\label{lemma: behavior of transmitted msg}
With the SED encoder described in \eqref{eq: MaxEJS encoder}, $\{U_{\thetai}(t)\}_{t=0}^{\infty}$ forms a submartingale with respect to the filtration $\F_t=\sigma\{Y^{t}\}$, satisfying
\begin{align}
\E[U_{\thetai}(t+1)|\F_t, \theta=i]\ge& U_{\thetai}(t)+C,\quad\text{if}\ U_{\thetai}(t)<0\label{eq: communication phase}\\
\E[U_{\thetai}(t+1)|\F_t, \theta=i]=& U_{\thetai}(t)+C_1,\quad\text{if}\ U_{\thetai}(t)\ge0\label{eq: confirmation phase}\\
|U_{\thetai}(t+1)-U_{\thetai}(t)|\le&C_2. \label{eq: bounded step size}
\end{align}
\end{lemma}
\begin{IEEEproof}
See Appendix \ref{appendix: proof of Lemma 1}.
\end{IEEEproof}

\begin{remark}
Lemma \ref{lemma: behavior of transmitted msg} characterizes the fundamental behavior of the transmitted message $\theta$. In particular, \eqref{eq: communication phase} and \eqref{eq: confirmation phase}  capture the dynamics of the transmitted message $\theta$ in communication and confirmation phases, respectively. Building on their analysis, we also prove the following result that enables us to upper bound $\E[\tau_i(1/2)|\theta=i]$.
\end{remark}

\begin{lemma}\label{lemma: the average step size}
With the SED encoder in \eqref{eq: MaxEJS encoder}, the log-likelihood ratios $\{U_i(t)\}_{t=0}^\infty$ of the transmitted message $\theta=i\in\Omega$ satisfies,
\begin{align*}
C\le \E[U_i(t+1)-U_i(t)|\theta=i]\le C_1.
\end{align*}
\end{lemma}
\begin{proof}
See Appendix \ref{appendix: proof of average step size}.
\end{proof}

\begin{lemma}[Naghshvar \emph{et al.}, \cite{Naghshvar2015}]\label{lemma: two submartingales}
Assume that the sequence $\{\xi_t\}_{t=0}^\infty$ forms a submartingale with respect to a filtration $\{\F_t\}$. Furthermore, assume there exist positive constants $K_1,K_2,$ and $K_3$ such that
\begin{align*}
\E[\xi_{t+1}|\F_t]\ge& \xi_t+K_1,\quad \text{if } \xi_t<0\\
\E[\xi_{t+1}|\F_t]\ge& \xi_t+K_2,\quad \text{if } \xi_t\ge 0\\
|\xi_{t+1}-\xi_t|\le& K_3.
\end{align*}
Consider the stopping time $v=\min\{t: \xi_t\ge B\}$, $B>0$. Then we have
\begin{align}
\E[v]\le \frac{B-\xi_0}{K_2}+\xi_0\indicator_{\{\xi_0<0\}}\left(\frac{1}{K_2}-\frac{1}{K_1} \right)+\frac{3K^2_3}{K_1K_2}. \label{eq: stopping time bound}
\end{align}
\end{lemma}
In \cite{Naghshvar2015}, the authors did not provide a proof of the almost sure finiteness of the stopping time defined in Lemma \ref{lemma: two submartingales}. For completeness, we state this fact in the following lemma.

\begin{lemma}\label{lemma: a.s. finite of stopping time}
Let $\{\xi_t\}_{t=0}^\infty$ be a submaringale with respect to a filtration $\{\F_t\}$ satisfying the conditions in Lemma \ref{lemma: two submartingales}. Then the stopping time $v=\min\{t: \xi_t\ge B\}$, $B>0$, is a.s. finite.
\end{lemma}
\begin{proof}
See Appendix \ref{appendix: proof of a.s. finiteness}.
\end{proof}

The submartingales in Lemma \ref{lemma: behavior of transmitted msg} can be incorporated into Lemma \ref{lemma: two submartingales} by setting $\xi_t=U_{\thetai}(t), K_1=C, K_2=C_1, K_3=C_2$ and $B=\log\frac{1-\epsilon}{\epsilon}$. Thus, appealing to \eqref{eq: stopping time bound}, we obtain the following tightened upper bound of $\E[\tau]$ over Theorem \ref{theorem: Naghshvar et al}.

\begin{corollary}\label{corollary: Naghshvar et al}
The proposed scheme described in \eqref{eq: stopping rule}, \eqref{eq: estimate message}, and \eqref{eq: MaxEJS encoder} for BSC$(p)$, $0<p<1/2$, satisfies
\begin{align}
\E[\tau]\le\frac{\log M}{C}+\frac{\log\frac{1-\epsilon}{\epsilon}}{C_1}+\frac{3C_2^2}{CC_1}. \label{eq: corollary of Naghshvar et al.}
\end{align}
\end{corollary}

\begin{remark}
The constant term in \eqref{eq: corollary of Naghshvar et al.} is less than that of \eqref{eq: Naghshvar et al} and thus provides an improved bound. However, this bound is still loose enough that its corresponding achievable rate lies below Polyanskiy's achievable rate for a system limited to stop feedback.
\end{remark}

Following Polyanskiy \cite{Polyanskiy2011}, Williamson \emph{et al.} \cite{Williamson2015} derived the VLF upper bound on average blocklength for the BSC.
\begin{theorem}[Polyanskiy's VLF bound, \cite{Williamson2015}]\label{theorem: VLF lower bound}
For a given $\epsilon>0$ and positive integer $M$, there exists a stop-feedback VLF code for BSC$(p)$, with average blocklength satisfying
\begin{align}
\E[\tau]\le\frac{\log(M-1)}{C}+\frac{\log\frac{1}{\epsilon}}{C}+\frac{\log 2(1-p)}{C}.
\end{align}
\end{theorem}

\begin{figure*}[t]
\centering
\includegraphics[width=0.9\textwidth]{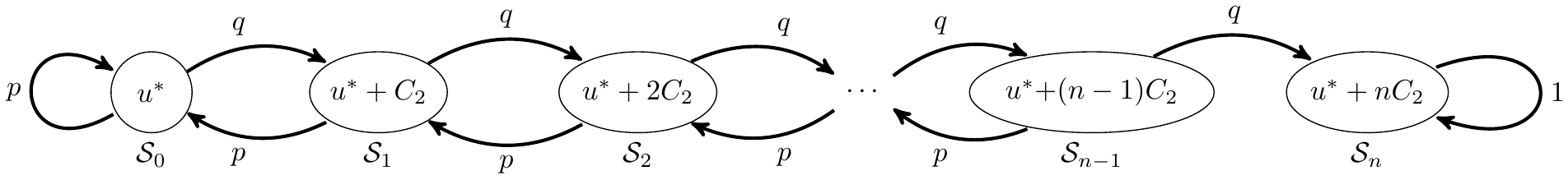}
\caption{An instance of the generalized Markov chain initiated at $U^*(t) = u^*$, where $u^*\in\St_0$ is some constant. The value inside the $i$-th circle is an element of state $\St_i$, $0\le i\le n$. }
\label{fig: generalized Markov chain}
\end{figure*}

\subsection{The Markovian Analysis: Proof of Theorem \ref{theorem: main results}}
\label{subsection: markovian analysis}

Consider the genie-aided decoder with the genie-aided stopping rule described in \eqref{eq: genie-aided decoder}. Clearly, $\tau\le \tau_{\thetai}(\epsilon)$ for any $\theta=i\in\Omega$, by definition. Thus,
\begin{align}
\E[\tau] =\E_{\theta}\big[\E[\tau|\theta=i]\big]\le\E_{\theta}\big[\E[\tau_i(\epsilon)|\theta=i]\big]=\E[\tau_i(\epsilon)|\theta=i]
\end{align}
where the last step follows in that the SED encoder does not reply on the choice of $\theta$. For any $\theta=i\in\Omega$, the SED encoder will yield the same average blocklength $\E[\tau_i(\epsilon)|\theta=i]$.

Next, we decompose $\E[\tau_{\thetai}(\epsilon)|\theta=i]$ as
\begin{align}
&\E[\tau_{\thetai}(\epsilon)|\theta=i]=\E[\tau_{\thetai}(1/2)+\tau_{\thetai}(\epsilon)-\tau_{\thetai}(1/2)|\theta=i]\notag\\
=&\E[\tau_{\thetai}(1/2)|\theta{=}i]+\E \big[\E[\tau_{\thetai}(\epsilon){-}\tau_{\thetai}(1/2)|\theta{=}i, U_{\thetai}(\tau_{\thetai}(1/2)){=}u]\big], \label{eq: decomposition}
\end{align}
where $\tau_{\thetai}(1/2)=\min\{t:\rho_{\thetai}(t)\ge1/2\}$ following \eqref{eq: genie-aided decoder} and $u$ represents the log-likelihood ratio of the transmitted message when $\rho_{\thetai}(t)$ crosses $1/2$ for the first time. By definition and Lemma \ref{lemma: behavior of transmitted msg}, $0\le u<C_2$. 

The decomposition in \eqref{eq: decomposition} provides a key insight on the average blocklength of the sequential transmission. It indicates that the overall average blocklength may be obtained as the sum of the expected time of first crossing of $1/2$ by $\rho_{\thetai}(t)$ and the expected time after the first crossing of $1/2$ until $\rho_{\thetai}(t)$ exceeds $1-\epsilon$. 

Appealing to Lemma \ref{lemma: two submartingales}, the expected time of first crossing of $1/2$ can be solved with submartingales. In order to bound the expected time after the first crossing of $1/2$ until $\rho_{\thetai}(t)$ exceeds $1-\epsilon$, we first show that $U_{\thetai}(t)$ forms a Markov chain when $U_{\thetai}(t)\ge0$. Thus, this time can be interpreted as the average of the conditional expected time-of-first passage from $U_{\thetai}(\tau_{\thetai}(1/2))=u$ to the destination $\log\frac{1-\epsilon}{\epsilon}$. However, one caveat is that this Markov chain should properly account for the \emph{fallback} from the confirmation phase into the communication phase and the subsequent return to the confirmation phase.

\begin{lemma}\label{lemma: first crossing of 1/2}
With the SED encoder in \eqref{eq: MaxEJS encoder}, the stopping time $\tau_i(1/2)$ of the transmitted message $\theta=i\in\Omega$ satisfies
\begin{align}
\E[\tau_{\thetai}(1/2)|\theta=i]<&\frac{\log M}{C}+\frac{C_1}{C}.
\end{align}
\end{lemma}
\begin{IEEEproof}
Let $\F_t=\sigma\{Y^t\}$ denote the history of receiver's knowledge up to time $t$. Consider $\eta_t=\frac{U_{\thetai}(t)}{C}-t$. First, we show that $\{\eta_t\}_{t=0}^\infty$ is also a submartingale. By Lemma \ref{lemma: two submartingales}, if $U_{\thetai}(t)<0$, we have
\begin{align}
\E[\eta_{t+1}|\F_t, \theta=i]=&\frac{\E[U_{\thetai}(t+1)|\F_t,\theta=i]}{C}-t-1\notag\\
	\ge&\frac{U_{\thetai}(t)+C}{C}-t-1\notag\\
	=&\eta_t.
\end{align}
If $U_{\thetai}(t)\ge0$, using the same argument with $C_1\ge C$, we can again show that $\E[\eta_{t+1}|\F_t, \theta=i]\ge \eta_t$. This implies that $\{\eta_t\}_{t=0}^{\infty}$ forms a submartingale. Let $T\triangleq \tau_i(1/2)$ be the shorthand notation for random variable $\tau_i(1/2)$. Since $T$ is a.s. finite, by Doob's optional stopping theorem \cite{Williams_1991},
\begin{align}
\E[\eta_0|\theta=i]\le \E[\eta_{T}|\theta=i],\label{eq: Doob's stopping theorem}
\end{align}
where
\begin{align}
\E[\eta_0|\theta=i]=&\frac{U_{\thetai}(0)}{C}=\frac{-\log(M-1)}{C},\label{eq: E_0}\\
\E[\eta_{T}|\theta=i]=&\frac{\E[U_i(T-1)|\theta=i]{+}\E[U_i(T){-}U_i(T-1)|\theta=i]}{C}\notag\\
	&-\E[T|\theta=i]\notag\\
	\le&\frac{0+C_1}{C}-\E[T|\theta=i].\label{eq: E_T}
\end{align}
where the last step follows from Lemma \ref{lemma: the average step size}. Substituting \eqref{eq: E_0} and \eqref{eq: E_T} into \eqref{eq: Doob's stopping theorem}, we have
\begin{align}
\E[T|\theta=i]<\frac{\log M}{C}+\frac{C_1}{C}.\label{eq: 32}
\end{align}
\end{IEEEproof}

\begin{lemma}\label{lemma: intermediate time}
With the SED encoder in \eqref{eq: MaxEJS encoder}, the difference between stopping times $\tau_i(\epsilon)$ and $\tau_i(1/2)$ of the transmitted message $\theta=i\in\Omega$ satisfies, for any $0\le u<C_2$,
\begin{align}
\E[\tau_{\thetai}(\epsilon)-&\tau_{\thetai}(1/2)\mid\theta=i, U_{\thetai}(\tau_{\thetai}(1/2))=u]\notag\\
&\le\frac{\left\lceil\frac{\log\frac{1-\epsilon}{\epsilon}}{C_2} \right\rceil C_2}{C_1}+\frac{pC_2}{C_1}\left(\frac{C_1+C_2}{C}-\frac{C_2}{C_1}\right).
\end{align}
\end{lemma}

\begin{IEEEproof}
The proof requires several steps. First, we show that if $\rho_{\thetai}(t)\ge1/2$ (or $U_{\thetai}(t)\ge0$), $U_{\thetai}(t)$ forms a Markov chain (or a random walk), which is given by Lemma \ref{lemma: Markov property}. Thus, $\E[\tau_{\thetai}(\epsilon)-\tau_{\thetai}(1/2)\mid\theta=i, U_{\thetai}(\tau_{\thetai}(1/2))=u]$ is equivalent to the expected time of first-passage from $u$ to $\log\frac{1-\epsilon}{\epsilon}$. However, such a Markov chain is still difficult to analyze because once the transmitted message $\theta$ falls back from $u$ and returns to the confirmation phase again, it may land at some other $u'$ different from $u$. Nevertheless, since the stopping time $\tau_i$ is a.s. finite, the transmitted message $\theta$ will return to the confirmation phase with probability $1$. This motivates the following \emph{generalized Markov chain}. 

\begin{definition}
Let $\St_0=[0, C_2)$ represent the set of all possible values of log-likelihood ratio $u$ when $\rho_{\thetai}(t)$ transitions from below $1/2$ to above $1/2$. Let $n\triangleq \lceil \log\frac{1-\epsilon}{\epsilon}/C_2 \rceil$. Let $\St_j=[jC_2, jC_2+C_2)$, $1\le j\le n$. The generalized Markov chain is defined as a sequence of states $\St_0, \St_1,\dots, \St_n$, satisfying
\begin{align*}
P(\St_{j+1}|\St_{j})=&P_{V|U}(u+C_2|u\in \St_{j-1})=q,\ 0\le j\le n-1\\
P(\St_{j-1}|\St_j)=&P_{V|U}(u-C_2|u\in \St_{j})=p, \quad 1\le j\le n,\\
P(\St_0|\St_0)=&P_{V|U}(u'\in \St_0|u\in \St_{0})=p,\\
P(\St_n|\St_n)=&P_{V|U}(u'\in\St_n|u\in \St_{n})=1.
\end{align*}
\end{definition}
The distinction between the generalized Markov chain and the regular Markov chain is that each state represents an interval rather than a single value. However, whenever $U_i(t)\ge0$, only one value in each state $\St_j$, $j=0,1,\dots,n$, remains active and those values can be readily determined from $U_i(t)$. Let $U^*(t)$ be the random variable denoting the value in state $\St_0$. Thus, $U^*(t)$ and $U(t)$ are related by
\begin{align}
    U^*(t)= \begin{cases}
        U_{\thetai}(t)-\left\lfloor\frac{U_{\thetai}(t)}{C_2} \right\rfloor C_2, & \text{if }\ U_{\thetai}(t)\ge0\\
        +\infty, & \text{otherwise}. \label{eq: initial position}
    \end{cases}
\end{align}
If $U^*(t)=u^*<\infty$, the active value in state $\St_j$ is given by $u^*+jC_2$. Furthermore, all active values remain constant as long as $ U_{\thetai}(t)\ge0$. Fig. \ref{fig: generalized Markov chain} illustrates an instance of the generalized Markov chain.

Let us consider the following position-invariant stopping rule on the generalized Markov chain
\begin{align}
    \tau^*_{\thetai}(\epsilon)=\min\left\{t: \left\lfloor\frac{U_{\thetai}(t)}{C_2} \right\rfloor\ge \left\lceil\frac{\log\frac{1-\epsilon}{\epsilon}}{C_2} \right\rceil \right\}. \label{eq: position-invariant stopping rule}
\end{align}
Thus, regardless of $U^*(t)$, the position-invariant stopping rule of \eqref{eq: position-invariant stopping rule} is achieved exactly when $U_{\thetai}(t)$ enters state $\St_n$ of the generalized Markov chain of Fig. \ref{fig: generalized Markov chain} for the first time.  In contrast, the stopping rule of \eqref{eq: genie-aided decoder} might be achieved either at state $\St_n$ or state $\St_{n-1}$, which complicates the analysis. 

More importantly, the position-invariant stopping rule is more stringent than the genie-aided stopping rule in that it yields an upper bound on $\tau_i(\epsilon)$, i.e.,
\begin{align}
    \tau\le \tau_{\thetai}(\epsilon)\le \tau_{\thetai}^*(\epsilon). \label{eq: upper bound on the genie}
\end{align}
This can be justified by the definition of $\tau_{\theta}(\epsilon)$ in \eqref{eq: genie-aided decoder} and that
\begin{align}
    \frac{U_{\thetai}(\tau_{\thetai}^*(\epsilon))}{C_2}\ge \left\lfloor\frac{U_{\thetai}(\tau_{\thetai}^*(\epsilon))}{C_2} \right\rfloor\ge \left\lceil\frac{\log\frac{1-\epsilon}{\epsilon}}{C_2}\right\rceil \ge \frac{\log\frac{1-\epsilon}{\epsilon}}{C_2}.
\end{align}
That is, $\rho_{\thetai}(\tau_{\thetai}^*(\epsilon))\ge 1-\epsilon$, which concludes that \eqref{eq: upper bound on the genie} holds.

 Let $V_i$ denote the expected time of first-passage from state $\St_i$ to state $\St_n$, $i=0,1,\dots,n-1$.  Thus, for any $0\le u^*<C_2$,
\begin{align}
    &\E[\tau_{\thetai}(\epsilon)-\tau_{\thetai}(1/2)\mid\theta=i, U_{\thetai}(\tau_{\thetai}(1/2))=u^*]\notag\\
    \le&\E[\tau_{\thetai}^*(\epsilon)-\tau_{\thetai}(1/2)\mid\theta=i, U_{\thetai}(\tau_{\thetai}(1/2))=u^*]=V_0.\label{eq: time-of-first passage inequality}
\end{align}

In Appendix \ref{appendix: time-of-first passage}, the time of first-passage analysis on the generalized Markov chain yields 
\begin{align}
V_0=\frac{n}{1-2p}+\frac{p}{1-2p}\left(1-\left(\frac{p}{1-p}\right)^n \right)(\Delta_0-\Delta^*_0)\label{eq: V_0}
\end{align}
as in \eqref{eq:V_0_differential}, where $\Delta^*_0$ is the expected self-loop time from state $\St_0$ to state $\St_0$ associated with a standard i.i.d. random walk as given by \eqref{eq:DeltaAll}, $\Delta_0$ is the actual expected self-loop time from state $\St_0$ to state $\St_0$, which is also the expected time it takes to fall back to the communication phase from state $\St_0$ and then return to state $\St_0$. Here, the
second term in \eqref{eq: V_0} is exactly the differential time between the actual process and the fictitious random walk, which reveals the fundamental mechanism of the constant term in the upper bound.  Using the same submartingale construction as in the proof of Lemma \ref{lemma: first crossing of 1/2}, we obtain
\begin{align}
\Delta_0\le&1+\frac{\E[U_{\thetai}(\tau_{\thetai}(1/2)-1)|\theta=i]+C_1-(u^*(t)-C_2)}{C}\notag\\
	\le&1+\frac{0+C_1+C_2}{C}. \label{eq: delta_0}
\end{align}
On the other hand, rewriting $\Delta^*_0$ in terms of $C_1, C_2$ yields
\begin{align}
\Delta^*_0=\frac{2-2p}{1-2p}=1+\frac{C_2}{C_1}.\label{eq: delta_0_star}
\end{align}
Therefore, combining \eqref{eq: V_0}, \eqref{eq: delta_0} and \eqref{eq: delta_0_star}, we have 
\begin{align}
V_0\le&\frac{n}{1-2p}+\frac{p}{1-2p}\left(\frac{C_1+C_2}{C}-\frac{C_2}{C_1}\right)\notag\\
	=&\frac{nC_2}{C_1}+\frac{pC_2}{C_1}\left(\frac{C_1+C_2}{C}-\frac{C_2}{C_1}\right). \label{eq: upper bound on V_0}
\end{align}
Finally, appealing to \eqref{eq: time-of-first passage inequality} and \eqref{eq: upper bound on V_0} concludes the proof.
\end{IEEEproof}

\begin{lemma}\label{lemma: Markov property}
The log-likelihood ratio $U_i(t)$ of the transmitted message $\theta=i\in\Omega$ satisfies
\begin{align}
P(U_{i}(t+1)=u+C_2|\theta=i, U_i(t)=u, u\ge0)=&q,\\
P(U_{i}(t+1)=u-C_2|\theta=i, U_i(t)=u, u\ge0)=&p. \label{eq:random walk}
\end{align}
\end{lemma}

\begin{IEEEproof}
If $U_{\thetai}(t)=u\ge0$, $\rho_{\thetai}(t)\ge1/2$. Then according to \eqref{eq: MaxEJS encoder}, the SED encoder will partition $\Omega$ into $S_0(t)=\{\thetai\}$ and $S_1(t)=\Omega\setminus\{\thetai\}$. Define the input probabilities
\begin{align}
\pi_x(t)=\sum_{j\in S_x(t)}\rho_j(t),\quad x\in\X. \label{eq: input probability}
\end{align}
Thus, $\pi_0(t)=\rho_{\thetai}(t)$, $\pi_1(t)=1-\rho_{\thetai}(t)$, and $X_t=0$. Therefore, the distribution of $Y_t$ is governed by law $P(Y|X=0)$. According to Bayes' rule in \eqref{eq: Bayes's rule},
\begin{align}
U_{\thetai}(t+1)=&\log\frac{\rho_{\thetai}(t+1)}{1-\rho_{\thetai}(t+1)}\notag\\
	=&\log\frac{\frac{\rho_{\thetai}(t)P(Y=y_t|X_t=0)}{\rho_{\thetai}(t)P(Y=y_t|X=0)+(1-\rho_{\thetai}(t))P(Y=y_t|X=1)}}{1-\frac{\rho_{\thetai}(t)P(Y=y_t|X_t=0)}{\rho_{\thetai}(t)P(Y=y_t|X=0)+(1-\rho_{\thetai}(t))P(Y=y_t|X=1)}}\notag\\
	=&\log\frac{\rho_{\thetai}(t)}{1-\rho_{\thetai}(t)}+\log\frac{P(Y=y_t|X_t=0)}{P(Y=y_t|X=1)}\notag\\
	=&\begin{cases}
	u+C_2, & y_t=0\ \text{with prob.}\ q\\
	u-C_2, & y_t=1\ \text{with prob.}\ p.
	\end{cases}
\end{align}
In the sequel, one can see that \eqref{eq: confirmation phase} is an immediate consequence of this Lemma.
\end{IEEEproof}

\section{Numerical Simulation}
\label{sec: simulation}

In this section, we consider the BSC with crossover probability $p=0.05$ and $\epsilon=10^{-3}$. Then, it can be calculated that
\begin{align}
C=0.7136,\ C_1=3.8231,\ C_2=4.2479.
\end{align}
One can verify that this setting satisfies the technical conditions in \cite{Naghshvar2015}. Thus, from \eqref{eq: Naghshvar et al} given by Naghshvar \emph{et al.},
\begin{align}
\E[\tau]\le\frac{\log M+\log\log M+3.32}{0.7136}+2.87+12702.89,
\end{align}
which turns out to be a loose bound.

\begin{figure}[t]
\centering
\includegraphics[width=0.48\textwidth]{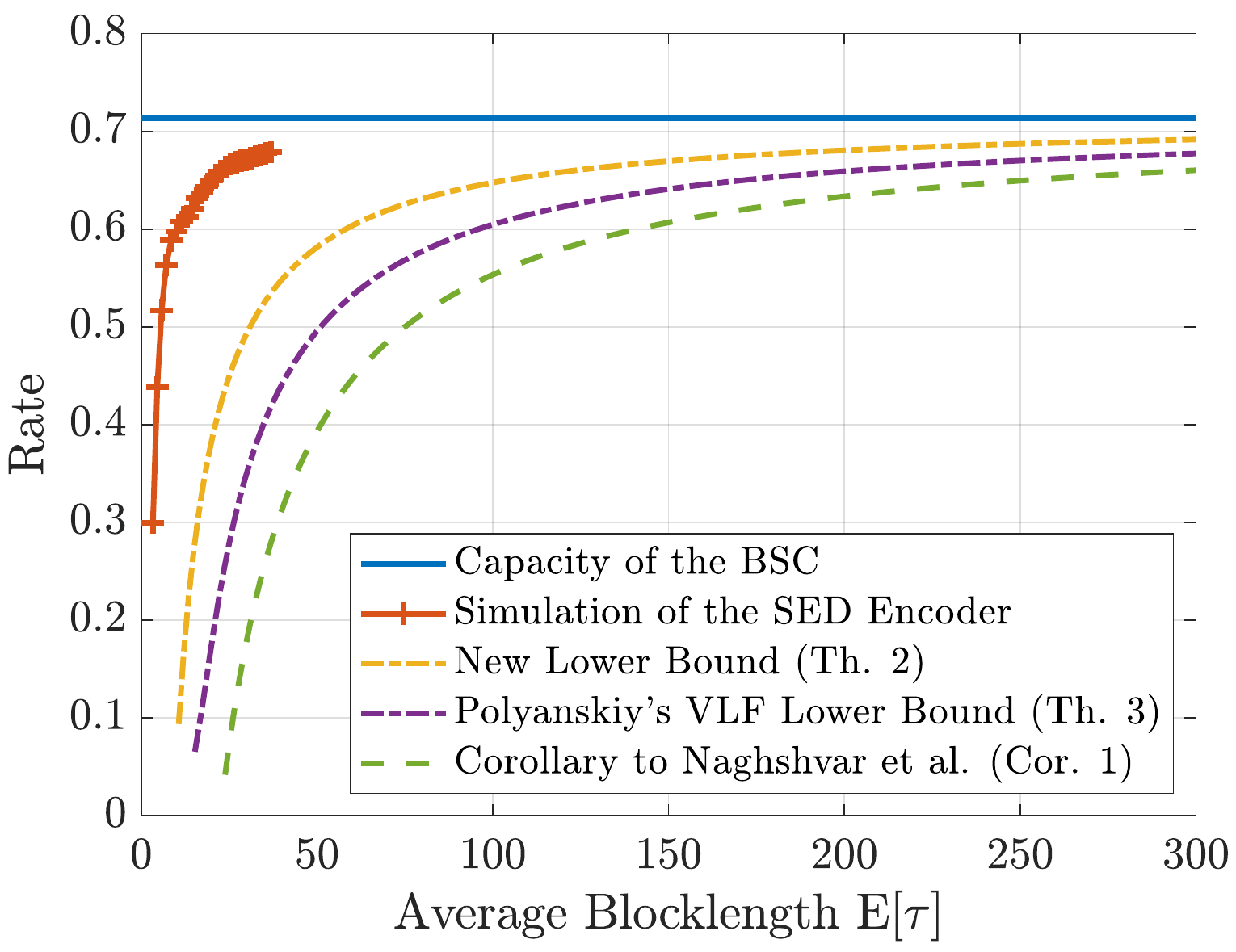}
\caption{The rate as a function of average blocklength over the BSC$(0.05)$ with full, noiseless feedback. $\epsilon=10^{-3}$. }
\label{fig: rate_vs_blocklength}
\end{figure}

The rate of a VLF code is given by
\begin{align}
R=\frac{\log M}{\E[\tau]}.
\end{align}

Fig. \ref{fig: rate_vs_blocklength} demonstrates the simulated rate performance of the SED encoder as a function of average blocklength $\E[\tau]$. Due to the exponential partitioning complexity, we were only able to simulate up to $k=25$, (or $M=2^{25}$). Since the upper bound on $\E[\tau]$ yields an achievability bound on rate, we also plot the achievability bounds given by Theorem \ref{theorem: main results}, Theorem \ref{theorem: VLF lower bound}, and Corollary \ref{corollary: Naghshvar et al}. One can see that our new bound exceeds the lower bound of Polyanskiy on achievable rate for a system limited to stop feedback, as would be expected for a system utilizing full, noiseless feedback. In contrast, Corollary \ref{corollary: Naghshvar et al} derived from Naghshvar \emph{et al.}'s submartingale result lies beneath Polyanskiy's VLF lower bound, indicating that it does not capture the actual performance of the SED encoder.

Indeed, we show analytically that our bound in Theorem \ref{theorem: main results} is tighter than that in Corollary \ref{corollary: Naghshvar et al} and is tighter than Polyanskiy's VLF bound in Theorem \ref{theorem: VLF lower bound} for moderately large crossover probability $p$.

\begin{theorem}\label{theorem: comparison}
For a given BSC$(p)$, $0<p<1/2$, the bound in Theorem \ref{theorem: main results} is strictly less than that in Corollary \ref{corollary: Naghshvar et al}. Furthermore, if $M\ge2$, $p\ge0.05$ and $\epsilon\le10^{-2}$, the bound in Theorem \ref{theorem: main results} is tighter than that in Theorem \ref{theorem: VLF lower bound}.
\end{theorem}

\begin{IEEEproof}
See Appendix \ref{appendix: proof of Theorem 4}.
\end{IEEEproof}

\section*{Acknowledgment}
The authors are grateful to Professor Tara Javidi whose talk at UCLA in 2012 inspired our initial interest in this work and whose comments about this work as it has developed have been critical to finally reaching the desired result.  Adam Williamson implemented the first simulation results in his doctoral work, which demonstrated that a smaller constant term than given in \cite{Naghshvar2015} might be possible with further analysis. Gourav Khadge helped implement the SED encoder for a wider range of average blocklengths.  Anonymous reviewers of this manuscript at various stages of its development provided invaluable constructive criticism.

\appendices

\section{Proof of Lemma \ref{lemma: behavior of transmitted msg} and Lemma \ref{lemma: the average step size}}

We briefly follow the proof as in \cite{Naghshvar2012}. The main proof requires the following lemma about the channel capacity. For brevity, we present this lemma here without proof. Interested readers can refer to \cite{Naghshvar2012} for further details.

\subsection{An Auxiliary Lemma}
\begin{lemma}[Naghshvar \emph{et al.}, \cite{Naghshvar2012}]\label{lemma: capacity lemma}
Let $P(Y|X)$ be a binary-input channel of positive capacity $C>0$. Let $P(X^*)$ be the capacity-achieving input distribution and $P(X)$ be an arbitrary input distribution for this channel. Also, let $P(Y^*)$ and $P(Y)$ be the output distributions induced by $P(X^*)$ and $P(X)$, respectively. Then, for any $x\in\X$ such that $P(X=x)\le P(X^*=x)$,
\begin{align*}
D\big(P(Y|X=x)\| P(Y) \big)\ge D\big(P(Y|X=x)\| P(Y^*) \big)=C.
\end{align*}
\end{lemma}

\subsection{Main Proof of Lemma \ref{lemma: behavior of transmitted msg} }
\label{appendix: proof of Lemma 1}
We first show \eqref{eq: communication phase} and \eqref{eq: confirmation phase} in Lemma \ref{lemma: behavior of transmitted msg}. Let $\theta=i\in S_{x_i}(t)$ be fixed, where $x_i\in\X$ is the channel input at time $t+1$. Define the extrinsic probabilities for the transmitted message $\theta=i$ as
\begin{align}
    \tilde{\pi}_{x,\thetai}(t)=\begin{cases}
        \frac{\pi_x(t)-\rho_{\thetai}(t)}{1-\rho_{\thetai}(t)}, & \text{if }\ i\in S_x(t)\\
        \frac{\pi_x(t)}{1-\rho_{\thetai}(t)}, & \text{if }\ i\notin S_x(t)
    \end{cases}\quad \forall x\in\X, \label{eq: extrinsic probability}
\end{align}
where $\pi_x(t)$ is defined in \eqref{eq: input probability}. Thus, $\sum_{x\in\X}\tilde{\pi}_{x,i}(t)=1$. Since $X_{t+1}=x_i$, $Y_{t+1}$ is distributed according to law $P(Y|X=x_i)$. Thus, we have
\begin{align}
    &\E[U_{\thetai}(t+1)-U_{\thetai}(t)|\F_t, \theta=i]\notag\\
    &=\E\left[\log\frac{\rho_{\thetai}(t+1)}{1-\rho_{\thetai}(t+1)}-\log\frac{\rho_{\thetai}(t)}{1-\rho_{\thetai}(t)}\middle |\F_t, \theta=i \right]\notag\\
    &=\sum_{y\in\Y}P(Y=y|X=x_i)\notag\\
    &\phantom{==}\cdot\left(\log\frac{\frac{\rho_{\thetai}(t)P(Y=y|X=x_i)}{\sum_{x\in\X}\pi_x(t)P(Y=y|X=x)}}{1-\frac{\rho_{\thetai}(t)P(Y=y|X=x_i)}{\sum_{x\in\X}\pi_x(t)P(Y=y|X=x)}}-\log\frac{\rho_{\thetai}(t)}{1-\rho_{\thetai}(t)} \right)\notag\\
    &=\sum_{y\in\Y}P(Y=y|X=x_i)\notag\\
    &\phantom{==}\cdot\left(\log\frac{P(Y=y|X=x_i)}{\sum_{x\in\X}\tilde{\pi}_{x,i}(t)P(Y=y|X=x)} \right)\notag\\
    &=D\big(P(Y|X=x_i)\| P(\tilde{Y}) \big), \label{eq: expected step size}
\end{align}
where $\tilde{Y}$ is the output induced by the channel $P(Y|X)$ for the input $\tilde{X}\sim P(\tilde{X}=x)=\tilde{\pi}_{x,i}(t)$.

When $U_{\thetai}(t)<0$, we further distinguish two cases. If $\theta=i\in S_1(t)$ and $x_i=1$:
\begin{align*}
    \tilde{\pi}_{1,i}(t)<0.5=P(X^*=1)
\end{align*}
because, by definition, $\pi_1(t)\le 0.5$ and $\tilde{\pi}_{1,i}(t)<\pi_1(t)$. Thus, by \eqref{eq: expected step size} and Lemma \ref{lemma: capacity lemma},
\begin{align}
    \E[U_{\thetai}(t+1)-U_{\thetai}(t)|\F_t, \theta=i]\ge C. \label{eq: expected step size greater than C}
\end{align}

If $\theta=i\in S_0(t)$ and $x_i=0$:
\begin{align*}
    \tilde{\pi}_{0,i}(t)\le 0.5=P(X^*=0)
\end{align*}
because, by the SED encoding rule in \eqref{eq: MaxEJS encoder} and the definition of extrinsic probabilities in \eqref{eq: extrinsic probability}, 
\begin{align*}
    \tilde{\pi}_{0,i}(t)=\frac{\pi_0(t)-\rho_{\thetai}(t)}{1-\rho_{\thetai}(t)}\le \frac{\pi_1(t)}{1-\rho_{\thetai}(t)}=\tilde{\pi}_{1,i}(t).
\end{align*}
By \eqref{eq: expected step size} and Lemma \ref{lemma: capacity lemma}, we again conclude \eqref{eq: expected step size greater than C}.

When $U_{\thetai}(t)\ge0$, then $\rho_{\thetai}(t)\ge0.5$ and by our encoding rule, $S_0(t)=\{i\}$ and $S_1=\Omega\setminus\{i\}$. Thus, $x_i=0$, $\pi_{0}(t)=\rho_{\thetai}(t)$, and $\tilde{\pi}_{0,i}(t)=0$. By \eqref{eq: expected step size},
\begin{align*}
    \E[U_{\thetai}(t+1)&-U_{\thetai}(t)|\F_t, \theta=i]\\
    &=D\big(P(Y|X=0)\| P(Y|X=1) \big)=C_1.
\end{align*}
Next, we show \eqref{eq: bounded step size} in Lemma \ref{lemma: behavior of transmitted msg}.
\begin{align*}
&|U_i(t+1)-U_i(t)|\\
=&\left|\log\frac{\rho_i(t+1)}{1-\rho_i(t+1)}-\log\frac{\rho_i(t)}{1-\rho_i(t)} \right|\\
=&\left|\log\frac{P(Y=y_{t+1}|X=e_{t+1}(i, y^t))}{\sum_{j\ne i}\frac{\rho_j(t)}{1-\rho_i(t)}P(Y=y_{t+1}|X=e_{t+1}(j, y^t)) } \right|\\
\le&\max_{y\in\Y}\log\frac{\max_{x\in\X}P(Y=y|X=x) }{\min_{x\in\X}P(Y=y|X=x) }\\
=&C_2.
\end{align*}

\subsection{Proof of Lemma \ref{lemma: the average step size} }
\label{appendix: proof of average step size}
Before we begin our proof, we appeal to the following auxiliary lemma regarding the KL divergence in \cite{Naghshvar2015}.
\begin{lemma}[Naghshvar \emph{et al.}, \cite{Naghshvar2015}]\label{lemma: decreasing KL divergence}
For any two distributions $P$ and $Q$ on a set $\Y$ and $\alpha\in[0,1]$, $D(P\| \alpha P+(1-\alpha)Q)$ is decreasing in $\alpha$.
\end{lemma}

Assume that $\theta=i\in S_{x_i}(t)$ for some $x_i\in\{0,1\}$. Let $\bar{x}_i = 1 - x_i$. In \eqref{eq: expected step size}, we have established that
\begin{align*}
\E[U_i(t+1)-U_i(t)|\F_t, \theta=i]=D\big(P(Y|X=x_i)\| P(\tilde{Y}) \big),
\end{align*}
where
\begin{align*}
P(\tilde{Y})=\tilde{\pi}_{x_i,i}(t)P(Y|X=x_i)+\tilde{\pi}_{\bar{x}_i,i}(t)P(Y|X=\bar{x}_i).
\end{align*}
Furthermore, we showed that $0\le \tilde{\pi}_{x_i,i}\le1/2$. Note that $\tilde{\pi}_{x_i,i}(t)+\tilde{\pi}_{\bar{x}_i,i}(t)=1$, therefore, $P(\tilde{Y})$ can be regarded as a mixture distribution between $P(Y|X=x_i)$ and $P(Y|X=\bar{x}_i)$. Appealing to Lemma \ref{lemma: decreasing KL divergence} and the BSC, we have
\begin{align}
&D\big(P(Y|X=x_i)\| P(\tilde{Y}) \big)\notag\\
&\le D(P(Y|X=x_i)\|P(Y|X=\bar{x}_i))=C_1 \label{eq: C1 upper bound}
\end{align}
Similarly, 
\begin{align}
&D\big(P(Y|X=x_i)\| P(\tilde{Y}) \big)\notag\\
&\ge D\left(P(Y|X=x_i)\|\frac12P(Y|X=x_i)+\frac12P(Y|X=\bar{x}_i)\right)\notag\\
&=C. \label{eq: C lower bound}
\end{align}
Since \eqref{eq: C1 upper bound} and \eqref{eq: C lower bound} hold for any $x_i$ and $y^t$ and note that $\E[U_i(t+1)-U_i(t)|\theta=i]=\E\big[\E[U_i(t+1)-U_i(t)|\F_t, \theta=i]\big]$, it follows that
\begin{align*}
C\le \E[U_i(t+1)-U_i(t)|\theta=i]\le& C_1.
\end{align*}

\section{Proof of Lemma \ref{lemma: a.s. finite of stopping time}}
\label{appendix: proof of a.s. finiteness}

To show that the stopping time $v=\min\{t: \xi_t\ge B\}$, $B>0$ is a.s. finite, we first recall Azuma's inequality for submartingales.
\begin{theorem}[Azuma's inequality]
If $\{\eta_t\}_{t=0}^\infty$ is a submartingale with respect to a filtration $\{\F_t\}$, satisfying $|\eta_{t+1}-\eta_t|\le K$, then for any $\epsilon>0$,
\begin{align}
\Pr\{\eta_t - \eta_0 \le -\epsilon \}\le\exp\left(\frac{-\epsilon^2}{2tK^2} \right).
\end{align}
\end{theorem}
Let $K=\min\{K_1, K_2\}$. Consider $\eta_t\triangleq \frac{\xi_t}{K}-t$. We show that $\{\eta_t\}_{t=0}^\infty$ is also a submartingale with respect to filtration $\{\F_t\}$.

If $\xi_t<0$, 
\begin{align*}
\E[\eta_{t+1}|\F_t] =&\frac{\E[\xi_{t+1}|\F_t]}{K}-t-1\\
	\ge&\frac{\xi_t+K_1 }{K}-t-1\\
	\ge&\eta_t.
\end{align*}
Similarly, we can show $\E[\eta_{t+1}|\F_t]\ge \xi_t$ for $\xi_t\ge0$. Hence, $\{\eta_t\}_{t=0}^\infty$ is a submartingale with respect to filtration $\{\F_t\}$. Furthermore, for any $t\ge0$,
\begin{align}
|\eta_{t+1} - \eta_t|=\left|\frac{\xi_{t+1}}{K} - \frac{\xi_t}{K}+1\right|\le\frac{|\xi_{t+1}-\xi_t|}{K}+1\le\frac{K_3}{K}+1.
\end{align}
Let $K_4\triangleq \frac{K_3}{K}+1$ for shorthand notation. Thus, appealing to Azuma's inequality,
\begin{align}
\Pr\{\xi_t\le (t-\epsilon)K+\xi_0 \}=&\Pr\left\{\frac{\xi_t}{K}-t-\frac{\xi_0}{K}\le -\epsilon \right\}\notag\\
	=&\Pr\{\eta_t-\eta_0\le\epsilon \}\notag\\
	\le&\exp\left(\frac{-\epsilon^2}{2tK_4^2} \right).
\end{align}
Equating $B = (t-\epsilon)K+\xi_0$, we have $\epsilon = t - \frac{B-\xi_0}{K}$, $t>\frac{B-\xi_0}{K}$. Hence,
\begin{align}
\Pr\{\xi_t \le B\} \le& \exp\left(\frac{-( t - \frac{B-\xi_0}{K})^2}{2tK_4^2} \right)\notag\\
	=&\exp\left(-\frac{t}{2K_4^2}+O(t^{-1})\right).
\end{align}
It follows that
\begin{align}
\lim_{t\to\infty} \Pr\{\xi_t\le B\}\le \lim_{t\to\infty}\exp\left(-\frac{t}{2K_4^2}+O(t^{-1})\right) = 0.\notag
\end{align}
This implies that
\begin{align}
\Pr\{v=\infty\} =& \lim_{t\to\infty}\Pr\Big(\bigcap_{i=1}^t\{\xi_i<B\} \Big)\notag\\
	\le&\lim_{t\to\infty}\Pr\{\xi_t<B\}=0.\notag
\end{align}
Namely, $\Pr\{v<\infty\}=1$.

\section{The Expected Time of First-Passage Analysis}
\label{appendix: time-of-first passage}

In this section we compute the expected time of first-passage $V_0$ for the generalized Markov chain, which is shown in Fig.  \ref{fig: generalized Markov chain}. Consider the general case of the Markov chain in Fig. \ref{fig: generalized Markov chain}, where the self-loop for state $\St_0$ has weight $\Delta_0$ and all other transitions in graph have weight $1$. Let $V_i$ be the expected time of first-passage from state $\St_i$ to state $\St_n$, $0\le i\le n-1$.  We wish to compute $V_0$.

This appendix computes $V_0$ by first simplifying the expected time-of-first-passage node equations into an expression involving only $V_0$ and $V_{n-1}$.  Characterizing the entire process to the left of $V_{n-1}$ as a self-loop with weight $\Delta_{n-1}$ yields an explicit expression for $V_{n-1}$.   This produces an expression for $V_0$ that naturally decomposes into the expected time of first-passage for a classic random walk plus an additional differential term.

\subsection{Simplifying node equations to involve only $V_0$ and $V_{n-1}$}
The node equations \cite{Gallagerbook} are as follows:
\begin{align}
V_{n-1} &= 1 + p V_{n-2} \label{eq:FirstNodeEqInc1}\\
V_{n-2} &= 1 + p V_{n-3} + q V_{n-1}\\
V_{n-3} &= 1 + p V_{n-4} + q V_{n-2}\\
V_{n-4} &= 1 + p V_{n-5} + q V_{n-3}\\
&~\vdots\notag\\
V_3 &= 1 + p V_2+q V_4\\
V_2 &= 1 + p V_1 + q V_3\\
V_1 &= 1 + p V_0 + q V_2\\
V_0 &= q + p V_0 + q V_1 + p \Delta_0\, . \label{eq:LastNodeEqInc1}
\end{align}
Summing the node equations described by  \eqref{eq:FirstNodeEqInc1}--\eqref{eq:LastNodeEqInc1} yields
\begin{equation*}
\sum_{i=0}^{n-1} V_i = n-1 + q + \sum_{i=1}^{n-2} V_i + q V_{n-1} + 2 p V_0 + p \Delta_0 \, ,
\end{equation*}
which simplifies to
\begin{equation*}
    V_0 + V_{n-1}=  n-1 + q + q V_{n-1} +  2 p V_0 + p \Delta_0 \, .
\end{equation*}
This yields 
\begin{equation}
V_0 = \frac{n-1 + q}{1-2p} +\frac{p}{1-2p} \left (\Delta_0 - V_{n-1}\right ) \, , \label{eq:V_0BeforeV_n-1KnownInc1}
\end{equation}
so that what remains to determine $V_0$ is to determine $V_{n-1}$.

\begin{figure}[t]
\centering
\includegraphics[width=3.3in]{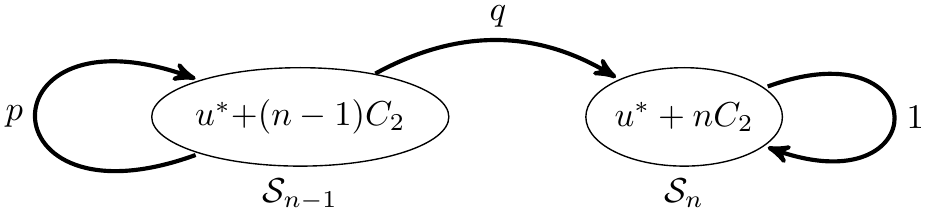}
\caption{The equivalent Markov chain from state $\St_{n-1}$ to state $\St_n$.}
\label{fig: V_n-1FSMS}
\end{figure}

\subsection{Finding $V_{n-1}$ using its left self-loop weight $\Delta_{n-1}$}
\label{sec:Deltafor1Inc}
We  determine $V_{n-1}$ in the general case for Fig.  \ref{fig: generalized Markov chain} by characterizing the entire process to the left of $\St_{n-1}$ as a the self-loop, as shown in Fig. \ref{fig: V_n-1FSMS}.

Let $\Delta_1$, be the the expected weight associated with the self-loop from state $\St_1$ that transitions to state $\St_0$ and then eventually returns to state $\St_1$. Regardless of what happens in state $\St_0$, at least two units of weight are accumulated by the initial transition to state $\St_0$ and the transition from state $\St_0$ back to state $\St_1$. With probability $p$ the weight-$\Delta_0$ self-loop is traversed at least once before state $\St_1$ is revisited, with probability $p^2$ weight-$\Delta_0$ self-loop is traversed a second time, and so on.  Thus the expected weight associated with traversing the zero-state self-loop with weight $\Delta_0$ is
\begin{equation*}
   \sum_{i=1}^{\infty} p^i \Delta_0 =  \left ( \frac{p}{1-p} \right )\Delta_0 \, .
\end{equation*}
Thus the expected weight associated with leaving state $\St_1$ by  traveling to state $\St_0$ and then returning to state $\St_1$ is
\begin{equation}
    \Delta_1  =  2 + \left ( \frac{p}{1-p} \right )\Delta_0 \, . \label{Delta1}
\end{equation}

\begin{remark} \label{remark:Delta_0^*}
We can use \eqref{Delta1} to find the left self-loop weight $\Delta_0^*$ for any state in a standard i.i.d. random walk where the state always transitions to the right with  probability $1-p$ and to the left with probability $p$.  Note that for such a random walk
$\Delta_{1}^* = \Delta_0^*$,  which implies from \eqref{Delta1} that 
\begin{equation*}
    \Delta_0^* = 2 + \left ( \frac{p}{1-p} \right )\Delta_0^* \, ,
\end{equation*}
so that 
\begin{equation}
      \Delta_0^* = \frac{2-2p}{1-2p} \, . \label{eq:DeltaAll}
\end{equation}
\end{remark}

Returning to the general case of Fig. \ref{fig: generalized Markov chain}, where $\Delta_0$ can have any value,  repeating the analysis that produced \eqref{Delta1} recursively yields
\begin{align}
    \Delta_{n-1} &= 2 \left[ \sum_{i=0}^{n-2} \left ( \frac{p}{1-p} \right ) ^i \right] + \left ( \frac{p}{1-p} \right ) ^{n-1}\Delta_0\notag\\
    &= 2 \left[\frac{1 - \left ( \frac{p}{1-p} \right )^{n-1}}{1 - \left ( \frac{p}{1-p} \right )}\right] + \left ( \frac{p}{1-p} \right ) ^{n-1}\Delta_0\, . \label{eq:Delta_n-1Inc1}
\end{align}
A time of first-passage analysis for $V_{n-1}$ using Fig. \ref{fig: V_n-1FSMS}  yields
\begin{align}
    V_{n-1} &= \left ( \frac{1}{1-p} \right ) \left ( p \Delta_{n-1} + 1-p \right )\notag\\
    &=\left (\frac{p }{1-p}\right ) \Delta_{n-1} +1\notag\\
    &= \left ( \frac{p}{1-p} \right ) ^{n}\Delta_0 + \frac{2p}{1-2p}\left ( 1 - \left ( \frac{p}{1-p} \right )^{n-1} \right ) + 1
    \, . \label{eq:V_n-1Inc1}
\end{align}

\subsection{Finding the general expression for $V_0$}
Substituting \eqref{eq:V_n-1Inc1}  into \eqref{eq:V_0BeforeV_n-1KnownInc1} yields
\begin{align*}
V_0 =& \frac{n-1 + q}{1-2p} +\frac{p}{1-2p} \left (\Delta_0 - V_{n-1}\right ) \\
= & \frac{n}{1-2p} + \frac{p \Delta_0}{1-2p} \left (1- \left ( \frac{p}{1-p} \right ) ^{n}\right )\\
&-\frac{2p^2}{(1-2p)^2}\left ( 1 - \left ( \frac{p}{1-p} \right )^{n-1} \right ) - \frac{2p}{1-2p}\\
= & \frac{n}{1-2p} + \frac{p \Delta_0}{1-2p} \left (1- \left ( \frac{p}{1-p} \right ) ^{n}\right )\\
&-\frac{2p -2p^2}{(1-2p)^2} + \frac{2p^2}{(1-2p)^2}\left ( \frac{p}{1-p} \right )^{n-1}\\
= & \frac{n}{1-2p} + \frac{p \Delta_0}{1-2p} \left (1- \left ( \frac{p}{1-p} \right ) ^{n}\right )\\&-\frac{2p -2p^2}{(1-2p)^2} + \frac{2p -2p^2}{(1-2p)^2}\left ( \frac{p}{1-p} \right )^{n}\\
= & \frac{n}{1-2p} \\
&+ \frac{p}{1-2p}\left ( 1- \left ( \frac{p}{1-p} \right ) ^{n} \right ) \left( \Delta_0 - \frac{2-2p}{1-2p}\right).
\end{align*}
Using the result in Remark \ref{remark:Delta_0^*}, this can be expressed as follows:
\begin{align}
V_0 =&\frac{n}{1-2p} 
+ \frac{p}{1-2p}\left ( 1- \left ( \frac{p}{1-p} \right ) ^{n} \right ) \left( \Delta_0 - \Delta_0^*\right). \label{eq:V_0_differential}
\end{align}
Note that when  $\Delta_0= \Delta_0^*$, \eqref{eq:V_0_differential}
simplifies to $ V_0=n/(1-2p)$ which is the expected time of first-passage $V_0^*$ for the standard random walk that was described in Remark \ref{remark:Delta_0^*}.

More generally, \eqref{eq:V_0_differential} expresses the expected time of first-passage as the sum of two terms.  The first term is equal to the expected time of first-passage for a standard random walk as described in Remark \ref{remark:Delta_0^*}, and the second term is a correction term we refer to as the ``differential time of first-passage''.   The differential time of first-passage depends on the difference between the self-loop weight $\Delta_0$ of the actual Markov chain under consideration and the self-loop weight $\Delta_0^*$ for a standard random walk as described in Remark \ref{remark:Delta_0^*}.

\section{Proof of Theorem \ref{theorem: comparison}}
\label{appendix: proof of Theorem 4}

Let $L_{\text{thm2}}(M,\epsilon)$, $L_{\text{thm3}}(M,\epsilon)$, and $L_{\text{cor1}}(M,\epsilon)$ denote the upper bounds in Theorem \ref{theorem: main results}, Theorem \ref{theorem: VLF lower bound}, and Corollary \ref{corollary: Naghshvar et al}, respectively. First, we show that $L_{\text{thm2}}(M,\epsilon)<L_{\text{cor1}}(M,\epsilon)$ for the BSC$(p)$, $0<p<1/2$,
\begin{align}
\phantom{}&L_{\text{thm2}}(M,\epsilon)\notag\\
\le&\frac{\log M}{C}+\frac{\log\frac{1-\epsilon}{\epsilon}}{C_1}+\frac{C_2}{C_1}+\frac{pC_2}{C_1}\left(\frac{C_1+C_2}{C}-\frac{C_2}{C_1}\right)+\frac{C_1}{C}\notag\\
    <&\frac{\log M}{C}+\frac{\log\frac{1-\epsilon}{\epsilon}}{C_1}+\frac{CC_2+C_1^2}{CC_1}+\frac{pC_2(C_1+C_2)}{CC_1}\notag\\
    \le&\frac{\log M}{C}+\frac{\log\frac{1-\epsilon}{\epsilon}}{C_1}+\frac{(2+2p)C_2^2}{CC_1}\notag\\
    \le&L_{\text{cor1}}(M,\epsilon).
\end{align}
Next, we show that if $M\ge2, \epsilon\le10^{-1}$, $L_{\text{thm2}}(M,\epsilon)<L_{\text{thm3}}(M,\epsilon)$. First,
\begin{align}
&L_{\text{thm3}}(M,\epsilon)-L_{\text{thm2}}(M,\epsilon)\notag\\
\ge&\frac{C_1-C}{CC_1}\log\frac{1}{\epsilon}+\frac{\log(1-p)}{C}-\frac{CC_2+C_1^2+pC_2(C_1+C_2)}{CC_1}. \label{eq: bound difference}
\end{align} 

For a given BSC$(p)$, $0<p<1/2$, we can view the difference in \eqref{eq: bound difference} as a function of $\epsilon$, i.e.,
\begin{align}
f(\epsilon)=\frac{C_1-C}{CC_1}\log\frac{1}{\epsilon}&+\frac{\log(1-p)}{C}\notag\\
 &-\frac{CC_2+C_1^2+pC_2(C_1+C_2)}{CC_1}.
\end{align}

Note that $f(\epsilon)$ is a monotonically decreasing function in $\epsilon$. Thus, solving for $f(\epsilon^*)=0$ yields
\begin{align}
\epsilon^*(p)=2^{-\frac{CC_2+C_1^2+pC_2(C_1+C_2)-C_1\log(1-p)}{C_1-C}}. \label{eq: threshold epsilon}
\end{align}
Since $\lim_{p\to0}\epsilon^*(p) = 0$. Therefore, $L_{\text{thm2}}(M,\epsilon)\le L_{\text{thm3}}(M,\epsilon)$ holds only for moderately large $p$ and small enough $\epsilon$. Numerical experiments show that if $p\ge0.05$ and $\epsilon\le10^{-2}<\epsilon^*(0.05)$, we have $L_{\text{thm2}}(M,\epsilon)\le L_{\text{thm3}}(M,\epsilon)$.

\bibliographystyle{IEEEtran}
\bibliography{IEEEabrv,references}

\end{document}